\documentclass[sigconf]{acmart}
\copyrightyear{2024}
\acmYear{2024}
\setcopyright{acmlicensed}
\acmConference[KDD '24] {Proceedings of the 30th ACM SIGKDD Conference on Knowledge Discovery and Data Mining }{August 25--29, 2024}{Barcelona, Spain.}
\acmBooktitle{Proceedings of the 30th ACM SIGKDD Conference on Knowledge Discovery and Data Mining (KDD '24), August 25--29, 2024, Barcelona, Spain}
\acmISBN{979-8-4007-0490-1/24/08}
\acmDOI{10.1145/3637528.3671725}

\usepackage{booktabs}
\usepackage{multirow}
\usepackage{makecell}
\usepackage{subfigure}
\usepackage[misc]{ifsym} 
\usepackage{balance}
\usepackage{listings}
\usepackage{color, xcolor}
\usepackage{algorithm2e}
\usepackage{tcolorbox}
\usepackage{amsmath}
\AtBeginDocument{%
	\providecommand\BibTeX{{%
			\normalfont B\kern-0.5em{\scshape i\kern-0.25em b}\kern-0.8em\TeX}}}

\acmSubmissionID{rtp0359}

\begin{document}
	\begin{sloppypar}
		
		\title[Multivariate Log-based Anomaly Detection for Distributed Database]{Multivariate Log-based Anomaly Detection \\ for Distributed Database}
		
		\author{Lingzhe Zhang}
		\affiliation{%
			\institution{Peking University}
			\city{Beijing}
			\country{China}}
		\orcid{0009-0005-9500-4489}
		\email{zhang.lingzhe@stu.pku.edu.cn}
		
		\author{Tong Jia$^{\ast}$}
		\thanks{*Corresponding author}
		\affiliation{%
			\institution{Peking University}
			\city{Beijing}
			\country{China}}
		\orcid{0000-0002-5946-9829}
		\email{jia.tong@pku.edu.cn}
		
		\author{Mengxi Jia}
		\affiliation{%
			\institution{Peking University}
			\city{Beijing}
			\country{China}}
		\orcid{0000-0002-0979-9803}
		\email{mxjia@pku.edu.cn}
		
		\author{Ying Li$^{\ast}$}
		\affiliation{%
			\institution{Peking University}
			\city{Beijing}
			\country{China}}
		\orcid{0000-0002-6278-2357}
		\email{li.ying@pku.edu.cn}
		
		\author{Yong Yang}
		\affiliation{%
			\institution{Peking University}
			\city{Beijing}
			\country{China}}
		\orcid{0000-0001-9667-2423}
		\email{yang.yong@pku.edu.cn}
		
		\author{Zhonghai Wu}
		\affiliation{%
			\institution{Peking University}
			\city{Beijing}
			\country{China}}
		\orcid{0000-0003-1268-836X}
		\email{wuzh@pku.edu.cn}
		
		\renewcommand{\shortauthors}{Lingzhe Zhang et al.}
		
		\begin{abstract}
			Distributed databases are fundamental infrastructures of today's large-scale software systems such as cloud systems. Detecting anomalies in distributed databases is essential for maintaining software availability. Existing approaches, predominantly developed using Loghub—a comprehensive collection of log datasets from various systems—lack datasets specifically tailored to distributed databases, which exhibit unique anomalies. Additionally, there's a notable absence of datasets encompassing multi-anomaly, multi-node logs. Consequently, models built upon these datasets, primarily designed for standalone systems, are inadequate for distributed databases, and the prevalent method of deeming an entire cluster anomalous based on irregularities in a single node leads to a high false-positive rate. This paper addresses the unique anomalies and multivariate nature of logs in distributed databases. We expose the first open-sourced, comprehensive dataset with multivariate logs from distributed databases. Utilizing this dataset, we conduct an extensive study to identify multiple database anomalies and to assess the effectiveness of state-of-the-art anomaly detection using multivariate log data. Our findings reveal that relying solely on logs from a single node is insufficient for accurate anomaly detection on distributed database. Leveraging these insights, we propose \textbf{MultiLog}, an innovative multivariate log-based anomaly detection approach tailored for distributed databases. Our experiments, based on this novel dataset, demonstrate MultiLog's superiority, outperforming existing state-of-the-art methods by approximately 12\%.
		\end{abstract}

		
		
		\maketitle
		
		\section{Introduction}
		
		The distributed databases, such as Google Spanner\cite{corbett2013spanner}, Alibaba OceanBase\cite{yang2022oceanbase}, PingCAP TiDB\cite{huang2020tidb}, and Apache IoTDB\cite{wang2020apache} have been widely used in cloud systems and is becoming a fundamental infrastructure to support the requirement for extremely high volume of data storage\cite{kang2022separation, zhang2024102224, zhang2024loadbalancing}.
		
		However, existing distributed databases suffer from frequent anomalies such as system failure, performance degradation, etc, and these anomalies often cause huge financial losses. For instance, Alibaba Cloud suffers from Intermittent Slow Queries (iSQs)\cite{ma2020diagnosing}, which result in billions of dollars in losses annually. Amazon also reports that every 0.1s of loading delay caused by database anomalies would cause an extra 1\% financial loss. As a result, detecting the anomalies of distributed databases and mitigating the affections of system anomalies are notoriously essential.
		
		As system logs meticulously track the states and significant events of actively running processes, they serve as a rich source for anomaly detection. Consequently, log-based anomaly detection has emerged as an effective method for ensuring software availability and has garnered extensive research attention. Most existing log-based anomaly detection models primarily utilize datasets from Loghub\cite{he2023loghub}, a comprehensive compilation of log datasets from a diverse range of systems. Among the most commonly utilized are systems such as standalone systems (e.g., BGL\cite{oliner2007supercomputers}, Thunderbird\cite{oliner2007supercomputers}, HPC\cite{makanju2009clustering}), distributed computing frameworks (like Hadoop\cite{lin2016log}, Spark), and distributed file systems (e.g., HDFS\cite{xu2009detecting}, Zookeeper.
		
		Leveraging Loghub, a variety of log-based anomaly detection models have been developed, broadly classified into two categories: supervised models\cite{chen2004failure, fronza2013failure, bodik2010fingerprinting, liang2007failure, zhang2019robust, yang2021semi, lu2018detecting} and unsupervised models\cite{babenko2009ava, du2017deeplog, jia2017logsed, yin2020improving, jia2021logflash,guo2021logbert, jia2022augmenting, kim2020automatic, meng2019loganomaly, landauer2023deep}. Supervised models, such as RobustLog\cite{zhang2019robust}, necessitate labeled data comprising both normal and abnormal instances to construct their predictive frameworks. In contrast, unsupervised models detect deviations relying solely on standard data. They are primarily split into deep neural network-based\cite{du2017deeplog, kim2020automatic, meng2019loganomaly, yin2020improving} and graph-based models\cite{babenko2009ava, jia2017logsed, jia2021logflash,guo2021logbert, jia2022augmenting, kim2020automatic}.
		
		Although these anomaly detection methods have demonstrated promising outcomes based on Loghub\cite{he2023loghub}, they still face the following practical challenges when applied to distributed databases:
		
		\begin{itemize}
			\item \textbf{Lack of Log Anomaly Datasets Collected from Distributed Databases:} Notably absent in Loghub are datasets specifically collected for distributed databases. This omission highlights a significant gap in the design, implementation, and testing of state-of-the-art models tailored for distributed databases. These databases are characterized by unique anomalies and features, necessitating the utlization of information from multiple dimensions, and many anomalies cannot be detected within the current node alone.
			
			\item \textbf{Absence of Datasets with Multi-Anomaly, Multi-Node Logs:} The majority of these datasets do not disclose the variety of anomaly types injected, often suggesting a limited scope of anomaly types. Even in datasets where the types are specified, like Hadoop\cite{lin2016log}, the range is limited (e.g., Machine down, Network disconnection, Disk full), lacking comprehensiveness\cite{landauer2023critical, huo2023autolog}. Additionally, these datasets typically comprise logs from single sources — either from standalone systems or single node of distributed systems, failing to capture the multi-node, interconnected nature of distributed databases.
			
			\item \textbf{Model Limitations for Distributed Databases:} Existing approaches are tailored for standalone systems, which do not align with the complexities of distributed databases. In these databases, to ensure data consistency, cluster nodes are categorized into Leaders and Followers, meaning different nodes can exhibit distinct information. Current methods, when used in distributed databases for what is termed as cluster anomaly detection, generally label the entire cluster as anomalous if any single node exhibits abnormal behavior. This approach can lead to a high rate of false positives, as it doesn't adequately consider the intricacies of distributed systems.
		\end{itemize}
		
		Recognizing this gap, we first construct a comprehensive dataset specifically designed for anomaly detection in distributed databases. Sourced from Apache IoTDB\cite{wang2020apache}, this dataset comprises 900 million records and spans 216 GB, making it the first dataset of this scale and specificity. We inject 11 types of anomalies, covering all categories reported in state-of-the-art research on distributed databases\cite{li2021opengauss, zhou2021dbmind, ma2020diagnosing}. It includes a wide range of anomalies, from resource-related issues to datatbase software faults. Moreover, to enable effective multivariate log-based anomaly detection, we meticulously collected logs from various nodes.
		
		We subsequently carry out an empirical study using this dataset to explore the characteristics of anomalies in distributed databases. Our study focuses on the performance of state-of-art anomaly detection models in identifying a spectrum of database anomalies, particularly in scenarios where logs from multiple database nodes are utilized. The findings of our study revealed a notable shortfall in these models – they are unable to reach peak performance when faced with database anomalies, and exhibit a high rate of false positives in detecting anomalies from multiple node logs. This inadequacy stems from the fact that different nodes within a database can present unique insights into anomalies, with certain anomalies being detectable only through the analysis of logs from multiple nodes.
		
		Building upon these insights, we introduce \textbf{MultiLog}, a multivariate log-based anomaly detection method for distributed database. In detail, we initially gather sequential, quantitative, and semantic information from the logs of each individual node within the database cluster. Subsequently, these results are processed through Standalone Estimation, where we utilize LSTM enhanced with self-attention to encode the data into a probability sequence, facilitating anomaly detection for each node. Finally, we introduce a Cluster Classifier that incorporates an AutoEncoder with a meta-classifier to standardize the probabilities across all nodes and identify the presence of any anomalies within the entire cluster.
		
		Our experiments demonstrate that MultiLog significantly surpasses current state-of-the-art methods. The evaluation results indicate that MultiLog excels in multi-node classification, achieving a remarkable improvement of approximately 12\% over existing methods. Furthermore, in the realm of anomaly detection within standalone database nodes, MultiLog notably boosts classification performance, recording an impressive enhancement of over 16\%. To summarize, our key contributions are as follows:
		
		\begin{itemize}
			\item We construct and open the first dataset for multivariate log-based anomaly detection on distributed databases, totaling 216 GB in size and comprising a total of 900 million records.
			\item We conduct a comprehensive study on database logs based on this dataset. Our study highlights that existing models face challenges in achieving optimal performance with randomly injected database anomalies. Notably, they exhibit a high rate of false positives when identifying anomalies across distributed database nodes.
			\item Inspired by the findings, we propose a multivariate log-based anomaly detection method for distributed database named \textbf{MultiLog}. This approach utilizes Standalone Estimation to encode sequential, quantitative, and semantic information from the logs of individual nodes, leveraging LSTM with self-attention for this purpose. Subsequently, MultiLog employs a Cluster Classifier, which integrates an AutoEncoder with a meta-classifier, to effectively classify anomalies within the database cluster.
			\item The effectiveness of MultiLog is confirmed based on our open dataset. Experiments show that MultiLog outperforms state-of-art methods in multi-nodes classification by approximately 12\% and improves classification effectiveness in standalone database nodes by over 16\%.
		\end{itemize}
		
		\begin{figure}[htbp]
			\centering
			\includegraphics[width=1\linewidth]{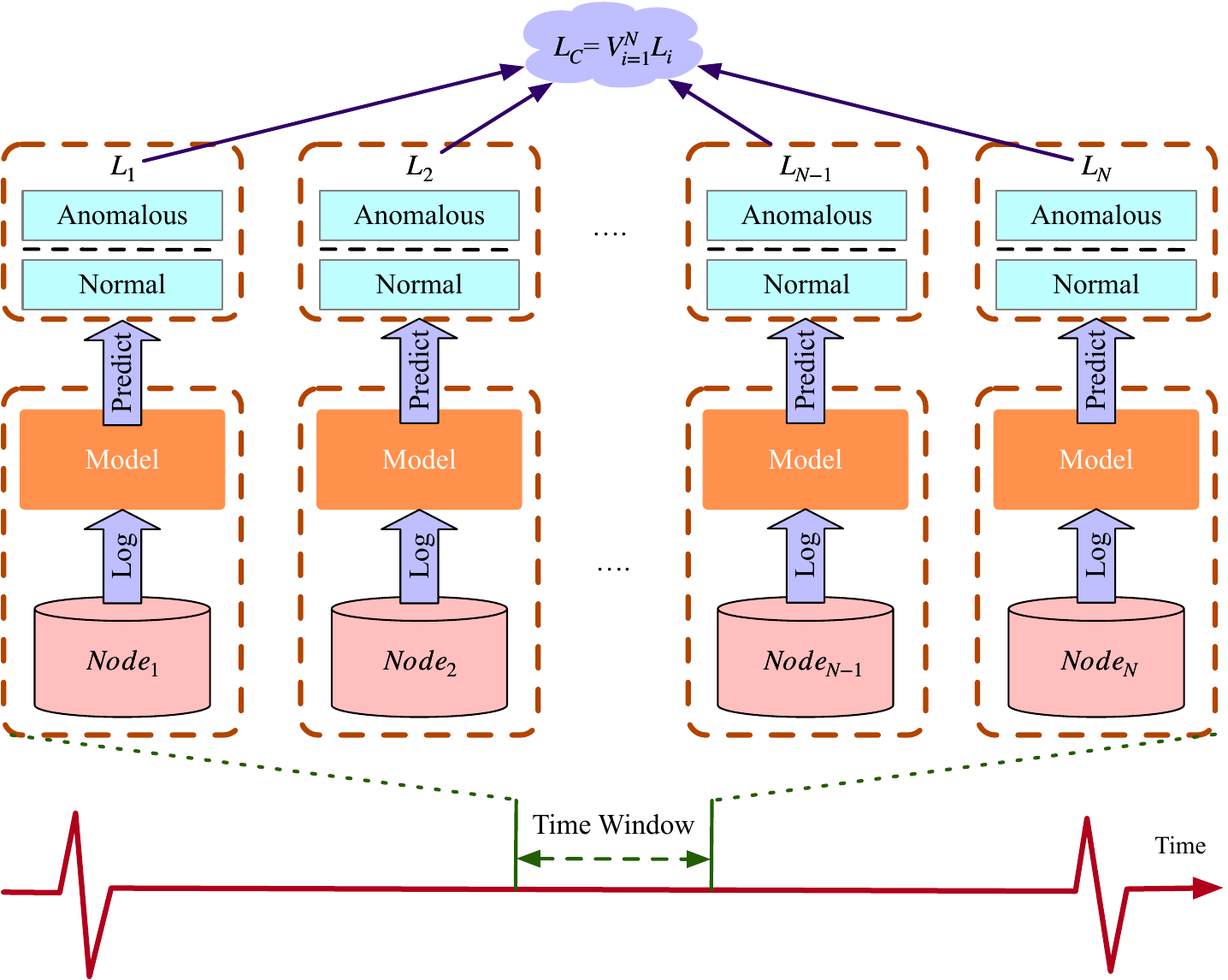}
			\caption{Application of Existing Models in a Distributed Context (Single-Point Classification)}
			\label{fig: standalone-detection}
		\end{figure}
		
		\section{Background}
		
		In this section, we outline the background of the log-based anomaly detection problem and clarify the typical application of these models within a distributed context.
		
		\subsection{Log-based Anomaly Detection}
		\label{sec: detection-background}
		
		A log sequence is composed of a series of log entries, arranged in the order they were output. An "event" represents an abstraction of a print statement in the source code, which appears in logs with varying parameter values across different executions. By using a set of invariant keywords and parameters (with parameters indicated by a placeholder "*"), an event can encapsulate multiple log entries. An event sequence is a sequence of log events, each corresponding one-to-one with log entries in a log sequence.
		
		We summarize the overall process of log-based anomaly detection in three steps: log parsing, log grouping, and model training. Log parsing extracts log events from system logs, and log grouping transforms the log sequence into an event sequence. Model training focuses on capturing the characteristics of event sequences and constructing models for automated recognition. In this paper, we apply Drain3\cite{he2017drain} for log parsing, a tool implemented by Logparser\cite{zhu2019tools, he2016evaluation}. For log grouping, we adopt a time-window based approach\cite{le2022log}. Additionally, our study utilizes three state-of-the-art models: RobustLog\cite{zhang2019robust}, LogAnomaly\cite{meng2019loganomaly}, and PLELog\cite{yang2021semi}.
		
		\subsection{Application of Anomaly Detection Models in a Distributed Context}
		
		When implementing the aforementioned log-based anomaly detection models in a distributed context, the prevailing approach is to assume the entire system is experiencing an anomaly if any single node within the cluster exhibits abnormal behavior.
		
		As illustrated in Figure~\ref{fig: standalone-detection}, consider a distributed cluster with $N$ nodes ($Node_{1}$,$Node_{2}$,...,$Node_{N-1}$,$Node_{N}$). At each time window $T$, a log-based anomaly detection model is applied to each $Node_{i}$ to predict whether the system is normal or anomalous, denoted as $L_{i} \in {0, 1}$. The label for the entire distributed cluster is then determined as $V_{i=1}^{N}L_{i}$. This implies that if any $L_{i}$ is anomalous within a given time window, the overall prediction for the cluster $L_{C}$ is classified as anomalous; otherwise, it is considered normal.
		
		\section{Construction of Dataset}
		
		In this section, we describe the construction of dataset. We conduct experiments using Apache IoTDB\cite{wang2020apache} v1.2.2 in a docker environment. Each docker cluster consists of one config node and one data node, with a configuration of 4 Intel(R) Xeon(R) Platinum 8260 CPUs at 2.40GHz, 16GB of DIMM RAM, 1.1TB NVMe disk, and running on openjdk11.
		
		For the generation of the data, we employ 4 write clients by default to insert data to four randomly selected nodes with each client assigned 100 threads for parallel insertion.
		
		Furthermore, to inject anomalies into the cluster we mainly utilize Chaos Mesh\cite{chang2015chaos, bjornberg2021cloud, mesh2021powerful}. It is an open-source cloud-native chaos engineering platform that provides a wide range of fault injection types, allowing users to simulate various types of anomalies that may occur in the real world. However, it is only suitable for simulating anomalies in docker environments. For anomalies related to workload and internal bugs, we employ dynamic read / write workload adjustment and database configuration modification approaches to inject.
		
		\subsection{Database Anomalies}
		
		We conduct an analysis of the definitions and classifications of anomalies in system and database work and discover that database anomalies can be classified into two categories\cite{wu2021microdiag, jeyakumar2019explainit, chen2014causeinfer, liu2021microhecl, liu2019fluxrank, bodik2010fingerprinting, zhou2018fault, ma2020diagnosing, yoon2016dbsherlock, li2021opengauss, zhou2021dbmind}: (1) anomalies caused by the system and (2) anomalies generated by the database itself.
		
		\begin{table*}[htbp]
			\centering
			\caption{Distributed Database Anomalies}
			\label{tab: anomalies}
			\begin{tabular}{cccc}
				\toprule
				No. & Anomaly & Cause Type & Description\\
				\midrule
				1 & CPU Saturation & System & The CPU computing resources exhaust.\\
				2 & IO Saturation & System & The I/O bandwidth is heavily occupied.\\
				3 & Memory Saturation & System & Insufficient memory resources.\\
				4 & Network Bandwidth Limited & System & The network bandwidth  between nodes is limited.\\
				5 & Network Partition Arise & System & Network partition occurs between nodes.\\
				6 & Machine Down & System & One server goes down when the applications are running.\\
				7 & Accompanying Slow Query & Database & Excessive query load.\\
				8 & Export Operations & Database & Backing up data to external source.\\
				9 & Import Operations & Database & Importing data from external source.\\
				10 & Resource-Intensive Compaction & Database & Compaction tasks consume a substantial amount of system resources.\\
				11 & Overly Frequent Disk Flushes & Database & The low interval of flush operations leads to frequent disk writes.\\
				\bottomrule
			\end{tabular}
		\end{table*}
		
		As outlined in Table~\ref{tab: anomalies}, we categorize 6 types of system-induced anomalies and 5 types of database-specific anomalies. Among these, Resource-Intensive Compaction\cite{o1996log, kang2022separation, sarkar2021constructing, zhang2021two, sarkar2023lsm, zhang2024102224} (No.10) is the most prevalent background task in LSM-based databases. Its impact on system CPU and memory resources varies based on configuration and can lead to write stall issues. Overly Frequent Disk Flushes (No.11) is a fault present in various types of databases. It pertains to the interaction between memory and disk in the database, significantly affecting the frequency of disk file creation and overall database performance.
		
		\subsection{Anomaly Injection}
		
		\begin{figure}[htbp]
			\centering
			\includegraphics[width=1\linewidth]{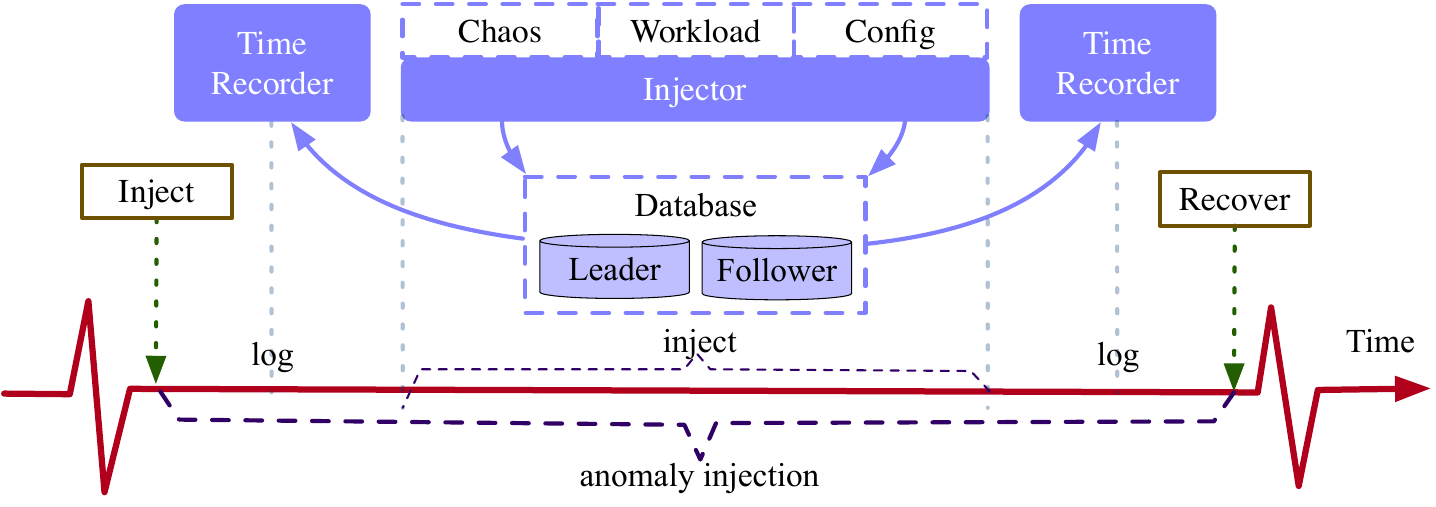}
			\caption{Architecture of Anomaly Injection}
			\label{fig: anomaly-injection-architecture}
		\end{figure}
		
		The main architecture of anomaly injection is demonstrated in Figure~\ref{fig: anomaly-injection-architecture}. During the experiment, the database initially operates normally for a set period. Subsequently, various types of anomalies are injected into the database. Before each injection, the current timestamp is recorded. The anomalies are then introduced and sustained for a specified duration. After the injection, the current timestamp is recorded again. The database is then allowed to resume normal operation for a certain period before the next anomaly is introduced. This process is repeated, with each cycle injecting a different type of anomaly, thus maintaining the database cluster in a cyclic state of normal - abnormal - normal - $\cdots$ - abnormal - normal.
		
		In the aforementioned architecture, the key component is the anomaly injector, which is responsible for injecting anomalies into the database. In the implementation, anomalies No. 1 to No. 6 utilize Chaos Mesh, anomalies No. 7 to No. 9 are accomplished by adjusting the read / write workload, and anomalies No. 10 to No. 11 are implemented through hot modifications to the database configuration.
		
		\subsection{Dataset Details}
		
		For the subsequent study, we employ the following various injection scenarios:
		
		\begin{itemize}
			\item \textbf{Single2Single:} This represents a \textbf{single} type of anomaly, injected into a \textbf{single} node. It encompasses two scenarios: the anomaly being injected either into the leader node or into a follower node.
			\item \textbf{Single2Multi:} a \textbf{single} type of anomaly, randomly injected into \textbf{multiple} nodes (including both leader node and follower nodes);
			\item \textbf{Multi2Single:} \textbf{multiple} types of anomalies, injected only into \textbf{single} node;
			\item \textbf{Multi2Multi:} \textbf{multiple} types of anomalies, randomly injected into \textbf{multiple} nodes (including both leader node and follower nodes).
		\end{itemize}
		
		\begin{table}[htbp]
			\centering
			\caption{Dataset Details}
			\label{tab: dataset-details}
			\begin{tabular}{ccccc}
				\toprule
				No. & Injection Type & \# Log Entries & Data Size & Labeled \\
				\midrule
				1 & Single2Single & 450,323,420 & 116.81 GB & $\checkmark$\\
				3 & Single2Multi & 245,215,090 & 47.35 GB & $\checkmark$\\
				4 & Multi2Single & 149,041,252 & 37.84 GB & $\checkmark$\\
				5 & Multi2Multi & 59,336,131 & 14.12 GB & $\checkmark$\\
				\bottomrule
			\end{tabular}
		\end{table}
		
		Based on the aforementioned process, the comprehensive statistics of the dataset are concluded in Table~\ref{tab: dataset-details}(the dataset will be disclosed upon acceptance). Overall, the dataset has a size of 216 GB and contains a total of 900 million records. The datasets are available in the MultiLog-Dataset repository, https://github.com/AIOps-LogDB/MultiLog-Dataset.
		
		\section{Empirical Study}
		
		In this section, we conduct a comprehensive study using the aforementioned dataset, with a focus on detecting database-specific anomalies, and identifying anomalies from multiple node logs.
		
		\subsection{Detecting Database-Specific Anomalies}
		
		We conduct an in-depth analysis to identify the relevant log information for database-specific anomalies. In the logs, valuable information that can be extracted includes sequential data, quantitative data, and semantic data. These respectively represent the sequence information, frequency information, and semantic information contained in the logs.
		
		\begin{figure}[htbp]
			\centering
			\includegraphics[width=1\linewidth]{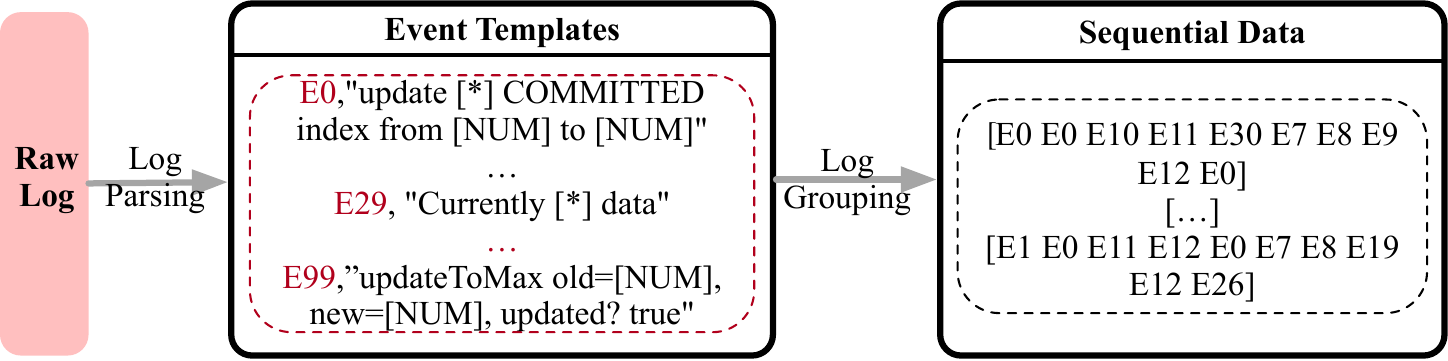}
			\caption{Sequential Data}
			\label{fig: sequential}
		\end{figure}
		
		As depicted in Figure~\ref{fig: sequential}, the predominant form of information is \textbf{sequential} data, specifically, the sequence of log entries. Raw logs undergo an initial parsing process into event templates, corresponding to each print statement in the code. This transformation simplifies the initially complex logs, rendering them as structured sequential information. Following this, a fixed-size window is applied to convert the obtained sequential information into organized windows, resulting in the final sequential data. This type of information proves particularly valuable in the detection of anomalies such as No.8 and No.9 export and import operations, as these operations often generate logs resembling 'Currently [*] data.
		
		\begin{figure}[htbp]
			\centering
			\includegraphics[width=1\linewidth]{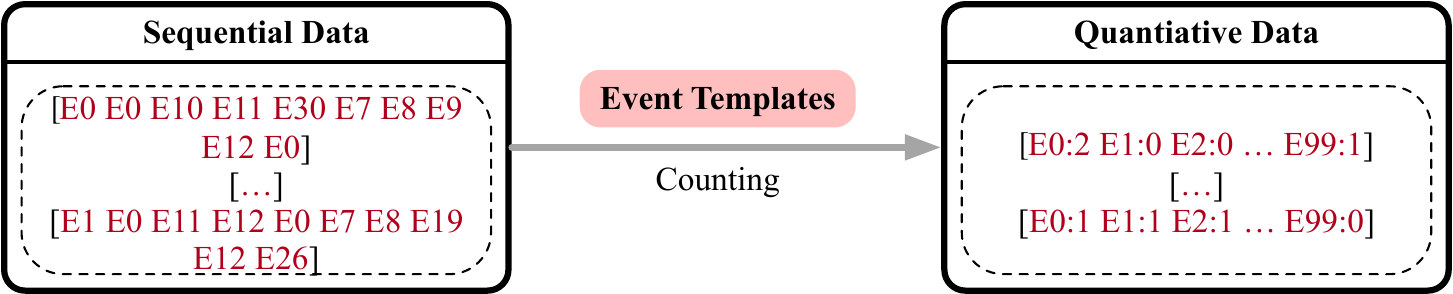}
			\caption{Quantitative Data}
			\label{fig: quantitative}
		\end{figure}
		
		The next aspect is \textbf{quantitative} information, as depicted in Figure~\ref{fig: quantitative}, which pertains to the frequency of each type of log entry within a window. Derived from the sequential data, this involves counting the occurrences of different event templates within a fixed-size window. This statistical approach is valuable for detecting anomalies such as No.11 Overly Frequent Disk Flushes, which are characterized by an excessive number of disk flush logs. 
		
		\begin{figure}[htbp]
			\centering
			\includegraphics[width=1\linewidth]{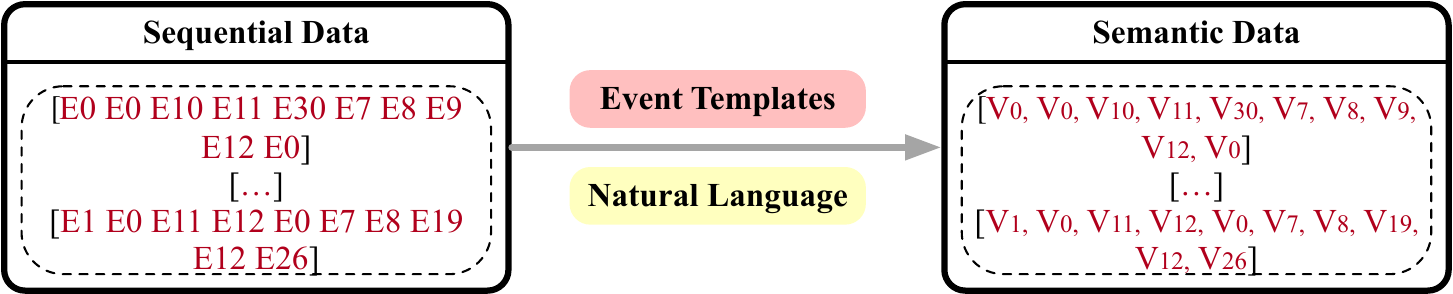}
			\caption{Semantic Data}
			\label{fig: semantic}
		\end{figure}
		
		The last aspect is \textbf{semantic} information, as depicted in Figure~\ref{fig: semantic}, which involves logs with inherent semantic content. Similar to the previous categories, this type of information is also derived from the sequential data. It transforms each event template into a semantic vector using sentence embedding methods. Then, the sequential data set in converted to a vector data set within a fixed-size window. This enrichment supplements the sequential data with meaningful semantic information. It proves valuable for leveraging internal database states in anomaly detection. For example, the presence of terms like 'error' or 'too high' can serve as significant indicators.
		
		\begin{center}
			\begin{tcolorbox}[colback=gray!10,
				colframe=black,
				width=\linewidth,
				arc=1mm, auto outer arc,
				boxrule=0.5pt,
				top=2pt, 
				bottom=2pt, 
				left=2pt,
				right=2pt
				]
				\textbf{Summary.} To classify database-specific anomalies, it is essential to gather information across multiple dimensions, including a mix of \textbf{sequential}, \textbf{quantitative}, and \textbf{semantic} information.
			\end{tcolorbox}
		\end{center}
		
		\subsection{Detecting Anomaly from Multiple Node Logs}
	
		We further analyze how individual nodes in a database cluster perceive anomalies when one is injected into a specific node.
		
		\begin{table}[h]
			\centering
			\caption{Evaluation Results for Each Node During Anomaly Injection (No.8 Export Operation) into $Node_{1}$}
			\label{tab: detecting-single-leader-export}
			\begin{tabular}{ccccc}
				\toprule
				\textbf{Model} & & Precision & Recall &  F1-Score \\
				\midrule
				\multirow{6}*{PLELog}
				& $Node_{1}$ & 39.68\% & 99.01\% & 56.66\%  \\
				~ & $Node_{2}$ & 34.35\% & 100.00\% & 51.14\% \\
				~ & $Node_{3}$ & 59.64\% & 98.02\% & 74.16\% \\
				~ & $Node_{4}$ & 31.17\% & 100.00\% & 47.53\% \\
				~ & $Node_{5}$ & 77.60\% & 96.04\% & 85.84\% \\
				~ & $Node_{6}$ & 89.42\% & 92.08\% & 90.73\% \\
				\bottomrule
				\multirow{6}*{RobustLog}
				& $Node_{1}$ & 55.81\% & 71.29\% & 62.61\% \\ 
				~ & $Node_{2}$ & 84.04\% & 98.75\% & 90.80\% \\
				~ & $Node_{3}$ & 24.44\% & 10.89\%& 15.07\% \\
				~ & $Node_{4}$ & 97.09\% & 99.01\% & 98.04\%  \\
				~ & $Node_{5}$ & 30.87\% & 100.00\% & 47.17\% \\
				~ & $Node_{6}$ & 100.00\% & 98.02\% & 99.00\% \\
				\midrule
			\end{tabular}
		\end{table}
		
		Firstly, \textbf{we find that the node directly affected by the anomaly faces challenges in detecting its own issue.} As shown in Figure~\ref{tab: detecting-single-leader-export}, when inject anomaly (No.8 Export Operation) into $Node_{1}$, PLELog results with a Precision of 39.68\%, Recall of 99.01\%, and F1-Score of 56.66\% and RobustLog results with a Precision of 55.81\%, Recall of 71.29\%, and F1-Score of 62.61\%. However, other nodes are able to detect this anomaly, with $Node_{6}$ achieving an F1-Score of 90.73\% in PLELog and 99.00\% in RobustLog. This situation arises because, although export-related anomalies generate explicit logs in the affected node, the patterns and frequency of these logs are relatively low, yet they significantly impact system performance. This leads to issues in information exchange with other nodes, making it possible to detect the anomaly from nodes other than the affected one. These experiments highlight the fact that relying solely on logs from a single node can cause the system to miss critical anomaly information, underscoring the importance of a multi-node perspective in anomaly detection.
		
		\begin{table}[htbp]
			\centering
			\caption{Evaluation Results for Each Node \& Cluster During Randomly Anomaly Injection (No.8 Export Operation) into Cluster}
			\label{tab: detecting-single-multi-export}
			\begin{tabular}{ccccc}
				\toprule
				\textbf{Model} & & Precision & Recall &  F1-Score \\
				\midrule
				\multirow{6}*{PLELog}
				& $Node_{1}$ & 72.55\% & 94.87\% & 82.22\%  \\
				~ & $Node_{2}$ & 40.21\% & 97.44\% & 56.93\% \\
				~ & $Node_{3}$ & 49.01\% & 94.87\% & 64.63\% \\
				~ & $Node_{4}$ & 93.98\% & 100.00\% & 96.89\% \\
				~ & $Node_{5}$ & 82.56\% & 91.03\% & 86.59\% \\
				~ & $Node_{6}$ & 59.35\% & 93.59\% & 72.64\% \\
				~ & Cluster & 74.82\% & 99.05\% & 85.25\% \\
				\midrule
				\multirow{6}*{RobustLog}
				& $Node_{1}$ & 70.83\% & 38.20\% & 49.64\%  \\
				~ & $Node_{2}$ & 91.55\% & 73.03\% & 81.25\% \\
				~ & $Node_{3}$ & 75.71\% & 59.55\%& 66.67\% \\
				~ & $Node_{4}$ & 85.07\% & 64.04\% & 73.08\% \\
				~ & $Node_{5}$ & 82.35\% & 15.73\% & 26.42\% \\
				~ & $Node_{6}$ & 87.88\% & 32.58\% & 47.54\% \\
				~ & Cluster & 65.57\% & 89.89\% & 75.83\% \\
				\bottomrule
			\end{tabular}
		\end{table}
	
		Therefore, when conducting cluster anomaly detection, it's crucial to utilize information from each node, which means employing Single-Point Classification is essential. To illustrate the significance of this approach, we conduct an experiment involving the random injection of anomaly No.8 into all nodes. \textbf{We find that the individual assessments of all nodes are not very effective.} As shown in Figure~\ref{tab: detecting-single-multi-export}, when anomaly No.8 is randomly injected. However, the use of Single-Point Classification enables us to achieve relatively better results for the entire cluster, although these results still fall short of the standards required for production environments in databases.
		
		\begin{table}[htbp]
			\centering
			\caption{Evaluation Results for Each Node \& Cluster During Anomaly Injection (No.7 Accompanying Slow Query) into $Node_{1}$}
			\label{tab: detecting-single-leader-query}
			\begin{tabular}{ccccc}
				\toprule
				\textbf{Model} & & Precision & Recall &  F1-Score \\
				\midrule
				\multirow{6}*{RobustLog}
				& $Node_{1}$ & 97.62\% & 96.47\% & 97.04\%  \\
				~ & $Node_{2}$ & 97.65\% & 97.65\% & 97.65\% \\
				~ & $Node_{3}$ & 92.22\% & 97.65\%& 94.86\% \\
				~ & $Node_{4}$ & 98.81\% & 97.65\% & 98.22\% \\
				~ & $Node_{5}$ & 92.13\% & 96.47\% & 94.25\% \\
				~ & $Node_{6}$ & 98.81\% & 97.65\% & 98.22\% \\
				~ & Cluster & 62.22\% & 100.00\% & 76.71\% \\
				\midrule
				\multirow{6}*{LogAnomaly}
				& $Node_{1}$ & 43.16\% & 97.62\% & 59.85\%  \\
				~ & $Node_{2}$ & 80.58\% & 98.81\% & 88.77\% \\
				~ & $Node_{3}$ & 93.33\% & 100.00\% & 96.55\% \\
				~ & $Node_{4}$ & 98.65\% & 86.90\% & 92.41\% \\
				~ & $Node_{5}$ & 94.94\% & 89.29\% & 92.02\% \\
				~ & $Node_{6}$ & 96.34\% & 94.05\% & 95.18\% \\
				~ & Cluster & 41.58\% & 100.00\% & 58.74\% \\
				\bottomrule
			\end{tabular}
		\end{table}
		
		Furthermore, we assess the overall state of the cluster for anomalies using the labels predicted by multiple nodes and employing the aforementioned Single-Point Classification method. We presents a detailed analysis of the evaluation results for each node and the cluster during the injection of anomaly No.7 (Accompanying Slow Query) into node $Node_{1}$. As shown in Figure~\ref{tab: detecting-single-leader-query}, it is noticeable that while the F1-Scores for anomaly detection at each individual node are generally high (mostly above 90\%), \textbf{the effectiveness drastically decreases when Single-Point Classification is applied.} This is evident in both RobustLog (with a 76.71\% F1-Score) and LogAnomaly (with a 58.74\% F1-Score), where a significant increase in false positives is observed (Reflected in 100\% Recall and lower Precision). However, the application of Single-Point Classification is essential, especially for anomalies such as No.8, where it is challenging to determine which node provides the correct label. 
		
		\begin{center}
			\begin{tcolorbox}[colback=gray!10,
				colframe=black,
				width=\linewidth,
				arc=1mm, auto outer arc,
				boxrule=0.5pt,
				top=2pt, 
				bottom=2pt, 
				left=2pt,
				right=2pt
				]
				\textbf{Summary.} In the context of cluster anomaly detection, relying solely on logs from a single node may result in missing many significant anomalies. However, using only Single-Point Classification to synthesize judgments from various nodes can lead to a high false positive rate.
			\end{tcolorbox}
		\end{center}
		
		\section{MultiLog: A Multivariate Log-Based Anomaly Detection Method For Distributed Database}
		
		\begin{figure*}[htbp]
			\centering
			\includegraphics[width=1\linewidth]{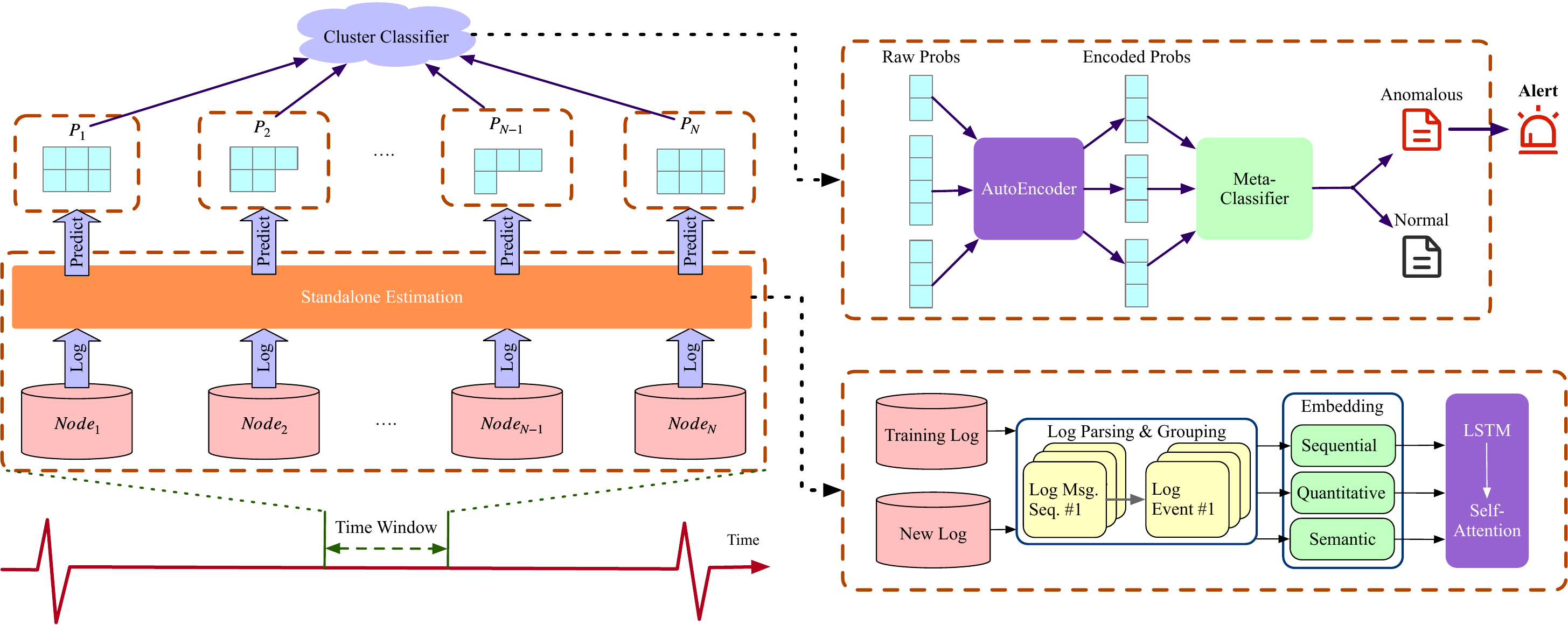}
			\caption{The Framework of MultiLog}
			\label{fig: multilog}
		\end{figure*}
		
		Our empirical study reveals that state-of-the-art models fail to achieve optimal detection results when faced with the injection of multiple anomalies. It also indicates that certain anomalies cannot be accurately classified by relying solely on logs from a single node, underscoring the need for a multivariate log approach in addressing cluster anomaly detection. However, both state-of-the-art models combined with Single-Point Classification method fall short in obtaining sufficiently accurate detection outcomes.
		
		Therefore, in this section, we introduce \textbf{MultiLog}, a multivariate log-based anomaly detection method specifically designed for distributed databases. Figure~\ref{fig: multilog} depicts the MultiLog framework. Unlike Single-Point Classification, which directly uses the predicted labels from multiple nodes to determine the cluster's anomaly label, MultiLog first employs Standalone Estimation to parse, embed, reduce dimensions, and compress logs from multiple nodes. This process ensures that, in each Time Window, each node outputs a variable-length probability list. Following this, the Cluster Classifier integrates these probability lists from each node to conclusively determine whether the current cluster is anomalous or normal.
		
		\subsection{Standalone Estimation}
		
		MultiLog transforms raw log sequences into three types of information: sequential, quantitative, and semantic, with the aim of preserving a comprehensive range of log data for later cluster classification. To capture long-term temporal dependencies and enhance crucial information, we further integrate self-attention. Summarily, Standalone Estimation encompasses five stages: log parsing \& grouping, sequential embedding, quantitative embedding, semantic embedding, and information enhancement.
		
		\subsubsection{Log Parsing \& Grouping.} As discussed in Section~\ref{sec: detection-background}, raw log messages are unstructured and include variable log parameters, presenting challenges to automated analysis. Thus, aligning with log-based anomaly detection practices, MultiLog employs log parsing to extract structured log events from these messages, thereby enabling more efficient analysis. Specifically, MultiLog utilizes the state-of-the-art Drain\cite{he2017drain} method for log parsing, which has been proven highly effective and efficient in existing studies\cite{zhang2023system}. After log parsing, the entirety of raw log messages in the current time window can be represented as $S = (s_{1}, s_{2}, ..., s_{M})$. Subsequently, for the convenience of further processing, we employ the fixed window method\cite{le2022log} to further segment $S$ into a series of fixed-length sequences $S_{j}$, each of length $M$.
		
		\subsubsection{Log Embedding.} After log grouping, each batch of raw log messages is parsed into log events, forming natural sequential patterns. The entirety of raw log messages can be represented as $E_{j} = (e_{(s_{1})}, e_{(s_{2})}, ..., e_{(s_{M})})$, where each event is represented as $e_{i}$. On this basis, we obtain quantitative patterns. To capture this aspect, we further perform the following transformation: For each group of log sequence $S_{j}$, we compute the count vector as $C_{j} = (c_{j} (e_{1}), c_{j} (e_{2}), ..., c_{j} (e_{n})$, where $c_j (e_{k})$ represents the frequency of $e_{k}$ in the event vector sequence $E_{j}$. Beyond sequential and quantitative patterns, an event sequence can also encapsulate semantic patterns. To extract the semantic information from log events, MultiLog treats each log event as a sentence in natural language. In detail, we can compute the semantic embedding of $E_{j}$ as $V_{j} = (v (e_{(s_{1})}), v (e_{(s_{2})}), ..., v (e_{(s_{M})}))$, where $v (e_{(s_{k})})$ represents the semantic embedding of log event $e_{(s_{k})}$. More detailed formula details are illustrated in Appendix~\ref{appendix: embedding}.
		
		\subsubsection{Information Enhancement.} For each pattern previously analyzed, we employ an LSTM with self-attention to amplify key information. Each embedding output— $E$, $C$, and $V$ —serves as the input for an LSTM. The LSTM's output is denoted as $H=[h_{1}, h_{2}, ..., h_{M}]$, where $h_{m}$ represents the hidden state at log event $m$, and $M$ is the size of the window. The self-attention mechanism, applied to this sequence of hidden states, is calculated as Equation~\ref{eq: self-attention}, with the scoring function defined as $score(h_{m},h_{M})=h_{m}^{\intercal} W h_{M}$, and $W$ as a learnable weight matrix.The context vector $c$, a weighted sum of the hidden states, is computed as $c=\sum_{m=1}^{M} \alpha_{m} h_{m}$. Subsequently, the enhanced output vector $EC$ for each pattern is obtained by concatenating the context vector $c$ with the last hidden state $h_{M}$, $EC = [c;h_{M}]$. Finally, these enhanced output vectors are concatenated and passed through a fully connected layer, represented mathematically as $p = FC([EC_{E};EC_{C};EC_{V}])$.
		
		\begin{equation}
			\alpha_{m} = \frac{exp(score(h_{m},h_{M}))}{\sum_{m'=1}^{M} exp(score(h_{m'},h_{M}))}
			\label{eq: self-attention}
		\end{equation}
		
		\subsection{Cluster Classifier}
		
		After acquiring the probabilities calculated by the Standalone Estimation from the database logs of each node, MultiLog employs the Cluster Classifier to predict the likelihood of the cluster being anomalous or normal. To standardize the length of the probabilities output by each node, an AutoEncoder is applied. Subsequently, MultiLog utilizes a meta-classifier to concatenate these standardized probabilities and predict the final outcome for the cluster in the current time window. In summary, the Cluster Classifier consists of two key steps: AutoEncoder for probability standardization and Meta-Classification for final cluster analysis.
		
		\subsubsection{AutoEncoder} Given that different nodes generate varying amounts of log data within the same time window, as previously described, we group logs based on a fixed length of $M$ and assign a probability $p \in \{0,1\}$ indicating the likelihood of anomaly for each group. Consequently, for a database with $N$ nodes, it can be inferred that in each time window, the probability list for node $Node_{i}$ is $P_{i}=[p_{1},p_{2},...,p{k_{i}}]$, where $p_{j}$ represents the final output for each group as mentioned earlier, and $k_{i}$ denotes the number of groups.
		
		\begin{subequations}
			\begin{align}
				& Z_{i} = f_{enc} (P_{i})
				\label{eq: encoder} \\
				& \hat{P_{i}} = f_{dec} (Z_{i})
				\label{eq: decoder}
			\end{align}
		\end{subequations}
	
		Subsequently, for the node outputs $\{P_{1}, P_{2}, ..., P_{N}\}$, MultiLog firstly apply the encoder function $f_{enc}$, which truncates or pads the input probability list $P_{i}$ to a fixed length $\beta$, and then maps it to a latent representation $Z_{i}$ of a fixed size $\mu$. This process is formulated in Equation~\ref{eq: encoder} and involves three linear layers with ReLU activation functions. The decoder function $d_{dec}$ then endeavors to reconstruct the original input from $Z_{i}$, as outlined in Equation~\ref{eq: decoder}, with the reconstructed probability list denoted as $\hat{P_{i}}$. To align the probability lists of each node, we utilize MSELoss to minimize the discrepancy between $P_{i}$ and $\hat{P_{i}}$, thereby achieving a standardized, dense representation $Z_{i}$ for each node's probability list.
		
		\subsubsection{Meta-Classification} After standardizing probabilities with the AutoEncoder, MultiLog employs Meta-Classification for the final assessment of the cluster's anomaly status. The latent representations from each node's standardized probability list, denoted as $\{Z_{1}, Z_{2}, ..., Z_{N}\}$, are concatenated to form a single vector as Equation~\ref{eq: concat-z}. This concatenated vector $Z$ is then fed into the meta-classifier. The meta-classifier, denoted as $f_{meta}$, is a neural network comprising one hidden layer and an output layer, the latter using a softmax activation function. The prediction of the overall anomaly status of the cluster is then formulated as in Equation~\ref{eq: meta-classification}. This process determines whether the cluster is classified as "anomalous" or "normal", based on the collective information from all nodes.
		
		\begin{subequations}
			\begin{align}
				& Z = [Z_{1};Z_{2};...;Z_{N}]
				\label{eq: concat-z} \\
				& P\{normal, anomalous\} = f_{meta} (Z)
				\label{eq: meta-classification}
			\end{align}
		\end{subequations}
	
	\begin{table*}[htbp]
		\centering
		\caption{F1-Score of Overall Evaluation}
		\label{tab: overall-evaluation}
		\begin{tabular}{c|ccc|ccc|ccc|c}
			\toprule
			\multirow{2}*{\textbf{Dataset}} & \multicolumn{3}{c}{PLELog} & \multicolumn{3}{c}{RobustLog} & \multicolumn{3}{c}{LogAnomaly} & MultiLog (our's) \\
			\\[-2.5ex]
			\cline{2-11}
			\\[-1.6ex]
			~ & \textbf{\textit{Single.}} & \textbf{\textit{Vote.}} & \textbf{\textit{Best.}} & \textbf{\textit{Single.}} & \textbf{\textit{Vote.}} & \textbf{\textit{Best.}} & \textbf{\textit{Single.}} & \textbf{\textit{Vote.}} & \textbf{\textit{Best.}} & \textbf{\textit{N.A.}}\\
			\midrule
			Single2Single & 43.98\% & 44.76\% & 80.95\% & 79.80\% & 85.18\% & 98.83\% & 93.40\% & 95.69\% & 98.80\% & 98.91\% \\
			Single2Multi & 42.62\% & 43.25\% & 79.49\% & 81.00\% & 82.75\% & 95.96\% & 93.19\% & 90.34\% & 96.45\% & 99.11\% \\
			Multi2Single & 38.94\% & 83.79\% & 51.02\% & 38.94\% & 98.16\% & 98.54\% & 95.96\% & 83.80\% & 99.56\% & 99.75\% \\
			Multi2Multi & 48.80\% & 33.12\% & 94.52\% & 66.63\% & 59.97\% & 97.42\% & 88.35\% & 55.97\% & 98.97\% & 99.82\% \\
			\bottomrule
		\end{tabular}
	\end{table*}
		
		\section{Experiment and Evaluation}
		
		In our experiments, we utilize the previously mentioned dataset to evaluate the effectiveness of MultiLog in comparison with state-of-the-art methods. Unless specified otherwise, we set the time window size to 5 seconds. For MultiLog, we configure the input size ($\beta$) of the AutoEncoder to be 128 and the output size ($\mu$) of the AutoEncoder to be 32.
		
		\subsection{Compared Approaches}
		
		Regarding the existing log-based anomaly detection approaches, we adopt PLELog\cite{yang2021semi}, RobustLog\cite{zhang2019robust} and LogAnomaly\cite{meng2019loganomaly}. In particular, we use the their public implementations\cite{github:logdeep, github:plelog} and determine their parameters by first reproducing the results in their corresponding studies.
		
		To conduct a more comprehensive comparison of MultiLog with these works, we additionally implement two additional methods for integrating logs from different nodes, beyond the Single-Point Classification approach.
		
		\textbf{Vote-Based Classification:} This method employs a voting system among nodes to make decisions, a concept often utilized in distributed consensus algorithms. It involves aggregating the decisions from individual nodes, where each node's prediction label ($L_{i} \in \{0, 1\}$) is considered a "vote". The final decision is determined based on the majority of votes, signifying that if $\sum_{i=1}^{N} L_i > \frac{N}{2}$ in a cluster with $N$ nodes, it is identified as anomalous; otherwise, it is considered normal.
		
		\textbf{Best-Node Classification:} This approach designates the node with the best overall performance as the deciding factor for cluster anomaly detection. While this method may not be feasible in real-time operation, it serves as a theoretical benchmark representing the optimal solution in certain scenarios. Essentially, it assumes that one node, due to its superior performance or accuracy in anomaly detection, can be the sole determinant of the cluster's state, thus providing insight into the best possible outcomes under ideal conditions.
		
		\subsection{Evaluation Results}
		
		We first conduct cluster anomaly detection based on state-of-art models: PLELog, RobustLog, and LogAnomaly, employing the Single-Point Classification, Vote-Based Classification, and Best-Node Classification methods as previously described, and compare them with our proposed MultiLog on our dataset.
		
		\begin{table}[htbp]
			\centering
			\caption{Experiment Results of Studied Approaches on Multi2Single and Multi2Multi Dataset}
			\label{tab: overall-evaluation-single}
			\begin{tabular}{c|c|ccc}
				\toprule
				\textbf{Dataset} & \textbf{Method} &\textbf{Precision} & \textbf{Recall} & \textbf{F1-Score} \\
				\midrule
				\multirow{4}*{Multi2Single} & PLELog & 24.18\% & 100.00\% & 38.94\% \\
				~ & RobustLog & 24.18\% & 100.00\% & 38.94\% \\
				~ & LogAnomaly & 95.88\% & 96.05\% & 95.96\% \\
				~ & MultiLog & 99.88\% & 99.62\% & 99.75\% \\
				\midrule
				\multirow{4}*{Multi2Multi} & PLELog & 34.52\% & 83.24\% & 48.80\% \\
				~ & RobustLog & 49.96\% & 100.00\% & 66.63\% \\
				~ & LogAnomaly & 72.23\% & 99.78\% & 83.80\% \\
				~ & MultiLog & 99.88\% & 99.75\% & 99.82\% \\ 
				\bottomrule
			\end{tabular}
		\end{table}
		
		As shown in Table~\ref{tab: overall-evaluation}, it is observable that for the Multi2Single dataset, RobustLog achieves the best result with Vote-Based Classification (F1-Score of 98.16\%), closely approaching the Best-Node Classification, which is impractical in real-world use, yet still fell short of MultiLog, which reaches 99.75\% on this dataset, surpassing the performance on PLELog, RobustLog, and LogAnomaly using Best-Node Classification. Furthermore, for the Multi2Multi dataset, the effectiveness of MultiLog far exceeds the best results of these three state-of-the-art models (LogAnomaly with Single-Point Classification), by approximately 11.5\%. This indicates that MultiLog, through the combination of Standalone Estimation and Cluster Classifier, can effectively perform anomaly detection in complex, multi-node scenarios with multiple anomalies. Moreover, even in simpler scenarios with a single type of anomaly (Single2Single, Single2Multi), MultiLog also achieved better results than the state-of-the-art models, with nearly 99\% F1-Score, surpassing the performance of the infeasible Best-Node Classification in state-of-the-art models. This demonstrates that MultiLog is highly effective even for single-anomaly scenarios.
		
		To delve deeper into the specifics under the Multi2Single and Multi2Multi datasets, we further conduct a detailed comparison between the state-of-the-art models (using Single-Point Classification, as this approach is commonly adopted in practical scenarios) and MultiLog. As depicted in Table~\ref{tab: overall-evaluation-single}, it is apparent that the superiority of MultiLog stems from its remarkably low false positive rate. Whether it’s PLELog, RobustLog, or LogAnomaly, their Recall rates are nearly 100\%, indicating their ability to almost always detect all anomalies. However, their Precision varies (ranging from 24.18\% for RobustLog to 72.23\% for LogAnomaly in the Multi2Multi dataset), whereas MultiLog’s Precision is also close to 100\%. In summary, MultiLog surpasses these state-of-the-art models chiefly due to its substantially lower rate of false positives, which, in practical production environments, can greatly reduce the workload of DBA (database administrator).
		
		To validate the effectiveness and stability of MultiLog, we conduct additional ablation experiments (Appendix~\ref{appendix: ablation}).
		
		\subsection{Limitations}
		
		Though our approach can significantly enhance anomaly detection performance with multivariate logs, there still exist several notable limitations. Firstly, our method is supervised. However, it can be adapted to unsupervised scenarios by integrating the Cluster Classifier with unsupervised models such as clustering. Secondly, due to increased network transmission and computational overhead, our approach operates less efficiently than state-of-the-art methods. In scenarios with very few instances of certain anomaly types, it may be relatively impractical. Thirdly, while our dataset includes a diverse range of injected anomalies compared to other datasets, it still deviates from real-world scenarios.
		
		\section{Related Work}
		
		\subsection{Log-Based Anomaly Detection and Diagnosis}
		
		Log analysis for fault detection and localization is a well-established research area\cite{du2017deeplog, jia2021logflash, jia2022augmenting, zhang2019robust, meng2019loganomaly, aclog, yang2021semi, chen2020logtransfer, han2021unsupervised, sui2023logkg, liu2022uniparser}. These methodologies typically involve extracting templates and key information from logs, followed by constructing models for anomaly detection and classification. There are mainly two types of models in this domain: graph-based and deep-learning models.
		
		Graph-based models leverage log events parsed from log files to create a graph-based representation. They detect conflicts and anomalies by comparing event sequences against this graph. For instance, LogFlash\cite{jia2021logflash} utilizes a real-time streaming process for log transitions, enhancing the speed of anomaly detection and diagnosis. HiLog\cite{jia2022augmenting} performs an empirical study on four anti-patterns that challenge the assumptions underlying anomaly detection and diagnosis models, proposing a human-in-the-loop approach to integrate human expertise into log-based anomaly detection. LogKG\cite{sui2023logkg} introduces a Failure-Oriented Log Representation (FOLR) method to extract failure-related patterns, using the OPTICS clustering method for anomaly diagnosis.
		
		Deep-learning models, conversely, use various neural networks to model sequences of log events. LogRobust\cite{zhang2019robust} applies Term Frequency-Inverse Document Frequency (TF-IDF) and word vectorization to convert log events into semantic vectors, thus improving the accuracy of anomaly detection. UniParser\cite{liu2022uniparser} employs a token encoder and a context encoder to learn patterns from log tokens and their adjacent contexts.
		
		\subsection{Anomaly Detection and Diagnosis for Database}
		
		Several anomaly detection and diagnosis approaches have been developed for databases\cite{liu2020fluxinfer, ma2020diagnosing, li2021opengauss, zhou2021dbmind}, particularly through the use of metrics data. FluxInfer\cite{liu2020fluxinfer} builds a weighted undirected dependency graph to illustrate the dependency relationships of anomalous monitoring data and employs a weighted PageRank algorithm for diagnosing online database anomalies. iSQUAD\cite{ma2020diagnosing} adopts the Bayesian Case Model to diagnose anomalies of intermittent slow queries (iSQs) in cloud databases, based on Key Performance Indicators (KPI) data. OpenGauss\cite{li2021opengauss, zhou2021dbmind}, an autonomous database system, implements an LSTM-based auto-encoder with an attention layer for system-level anomaly diagnosis, also leveraging features in metrics data. Sentinel\cite{glasbergen2020sentinel} constructs a fine-grained model of data system behavior through debug logs to assist DBAs in diagnosing anomalies.
		
		To the best of our knowledge, we are the first to construct a dataset specifically designed for log-based anomaly detection in databases. Distinct from prior research, our focus is on leveraging database logs instead of metric data. Additionally, our approach emphasizes the utilization of logs from distributed nodes, setting it apart from related work that concentrates on addressing anomaly detection issues in standalone nodes.
		
		\section{Conclusions}
		
		In this paper, we study the problem of multivariate log-based anomaly detection in distributed databases. Initially, we conduct a comprehensive study to evaluate state-of-the-art anomaly detection models specific to distributed databases. This study reveals that these models experience a decline in classification effectiveness when dealing with multiple database anomalies and exhibit a high false positive rate when using multivariate logs. Based on the study, we introduce MultiLog, a multivariate log-based anomaly detection approach for distributed database. MultiLog adeptly leverages multivariate logs from all nodes within the distributed database cluster, significantly enhancing the efficacy of anomaly detection. For evaluation, we release the first open-sourced comprehensive dataset featuring multivariate logs from distributed databases. Our experimental evaluations based on this dataset demonstrate the effectiveness of MultiLog.
		
		In future work, our research will concentrate on incorporating more characteristics of distributed databases for anomaly detection. Additionally, we will further explore anomaly diagnosis and root cause analysis for distributed databases based on log data.
		
		\section*{Acknowledgement}
		
		This work was supported by the National Key R\&D Research Fund of
		China (2021YFF0704202).
		
		\bibliographystyle{ACM-Reference-Format}
		\balance
		\bibliography{sample-base}
		
		\clearpage
		
		\appendix
		
		\section{Log Embedding}
		\label{appendix: embedding}
		
		\subsection{Sequential Embedding.} After log grouping, each batch of raw log messages is parsed into log events, forming natural sequential patterns. Specifically, each event is represented as a distinct vector $e_{i}$, and the entire collection of unique event vectors is denoted as $\omega = \{e_{1} , e_{2} , ..., e_{n}\}$. Consequently, the event vector sequence $S_{j}$ is transformed as Equation~\ref{eq: sequential}, where $E_{j}$ constitutes the sequential embedding of this particular group of log message sequences.
		
		\begin{equation}
			E_{j} = (e_{(s_{1})}, e_{(s_{2})}, ..., e_{(s_{M})})
			\label{eq: sequential}
		\end{equation}
		
		\subsection{Quantitative Embedding.} Beyond sequential patterns, an event sequence also exhibits quantitative patterns. Typically, normal program execution follows certain invariants, with specific quantitative relationships consistently maintained in logs under various inputs and workloads. For instance, each file opened during a process is eventually closed. As such, the number of logs indicating "create a memtable" should match those showing "flush a memtable to file" under normal conditions. These quantitative relationships within logs are indicative of standard program execution behaviors. Anomalies can be identified when new logs disrupt these invariants. To capture this aspect, we analyze the quantitative pattern of logs as follows: For each group of log sequence $S_{j}$, we compute the count vector as Equation~\ref{eq: quantitative}, where $c_j (e_{k})$ represents the frequency of $e_{k}$ in the event vector sequence $(s_{1}, s_{2}, ..., s_{m})$, and $e_{k} \in \omega$.
		
		\begin{equation}
			C_{j} = (c_{j} (e_{1}), c_{j} (e_{2}), ..., c_{j} (e_{n})
			\label{eq: quantitative}
		\end{equation}
		
		\subsection{Semantic Embedding.} Beyond sequential and quantitative patterns, an event sequence can also encapsulate semantic patterns. To extract the semantic information from log events, MultiLog treats each log event as a sentence in natural language. Given that log events are formulated by developers to record system statuses, most tokens in these events are English words with intrinsic semantics. However, log events also contain non-character tokens (like delimiters, operators, punctuation marks, and number digits) and composite tokens resulting from word concatenations (such as "NullPointerException") due to programming conventions. In alignment with existing research\cite{yang2021semi}, MultiLog initially preprocesses log events by eliminating non-character tokens and stop words, and decomposing composite tokens into individual words using Camel Case splitting. Following this, MultiLog utilizes pre-trained word vectors based on the Common Crawl Corpus with the FastText algorithm\cite{pennington2014glove}, which is adept at capturing the inherent relationships among words in natural language. This means each word in a processed log event is transformed into a $d$-dimensional vector (denoted as $v$) using the pre-trained word2vec model, where $d$ is set to 300 in FastText word vectors.
		
		\begin{subequations}
			\begin{align}
				& TF(w) = \frac{\#w}{\#W}
				\label{eq: tf} \\
				& IDF(w)=log(\frac{\#L}{\#L_{w}})
				\label{eq: idf}
			\end{align}
		\end{subequations}
		
		After transforming each word to a $d$-dimension vector via FastText word embedding, MultiLog further transforms a log event into a semantic vector by aggregating all the word vectors within it. For this aggregation, MultiLog adopts TF-IDF\cite{salton1988term}, taking into account the significance of each word. Term Frequency (TF) measures how frequently a word assesses the frequency of a word $w$ in a log event, calculated as Equation~\ref{eq: tf}, where $\#w$ represents the count of $w$ in the log event and $\#W$ represents the total word count in the log event. Inverse Document Frequency (IDF) evaluates the commonality or rarity of $w$ iacross all log events, computed as Equation~\ref{eq: idf}, where $\#L$ is the total number of log events, and $\#L_{w}$ is the number of log events containing $w$. The weight of a word (denoted as $\epsilon$) is determined by $TF \times IDF$. Ultimately, the semantic vector (denotes $V$) of a log event is derived by summing up all word vectors in the log event, weighted by their TF-IDF values, as $V=\frac{1}{W} \sum_{i=1}^{W} \epsilon_{i}$.
		
		\section{Effectiveness of Cluster Classifier}
		\label{appendix: ablation}
		
		Although the effectiveness of MultiLog is a combined result of the Standalone Estimation and Cluster Classifier, the latter plays a crucial role in utilizing information from multiple nodes. Therefore, to further analyze its impact, we conduct comparative validation experiments on both the Multi2Single and Multi2Multi datasets. For the probability lists resulting from Standalone Estimation, we employ both the  Single-Based Classification and Vote-Based Classification to judge whether the cluster is anomalous or normal.
		
		\begin{figure}[htbp]
			\centering
			\subfigure[Multi2Single]{
				\begin{minipage}{0.47\linewidth}
					\centering   
					\includegraphics[width=\textwidth]{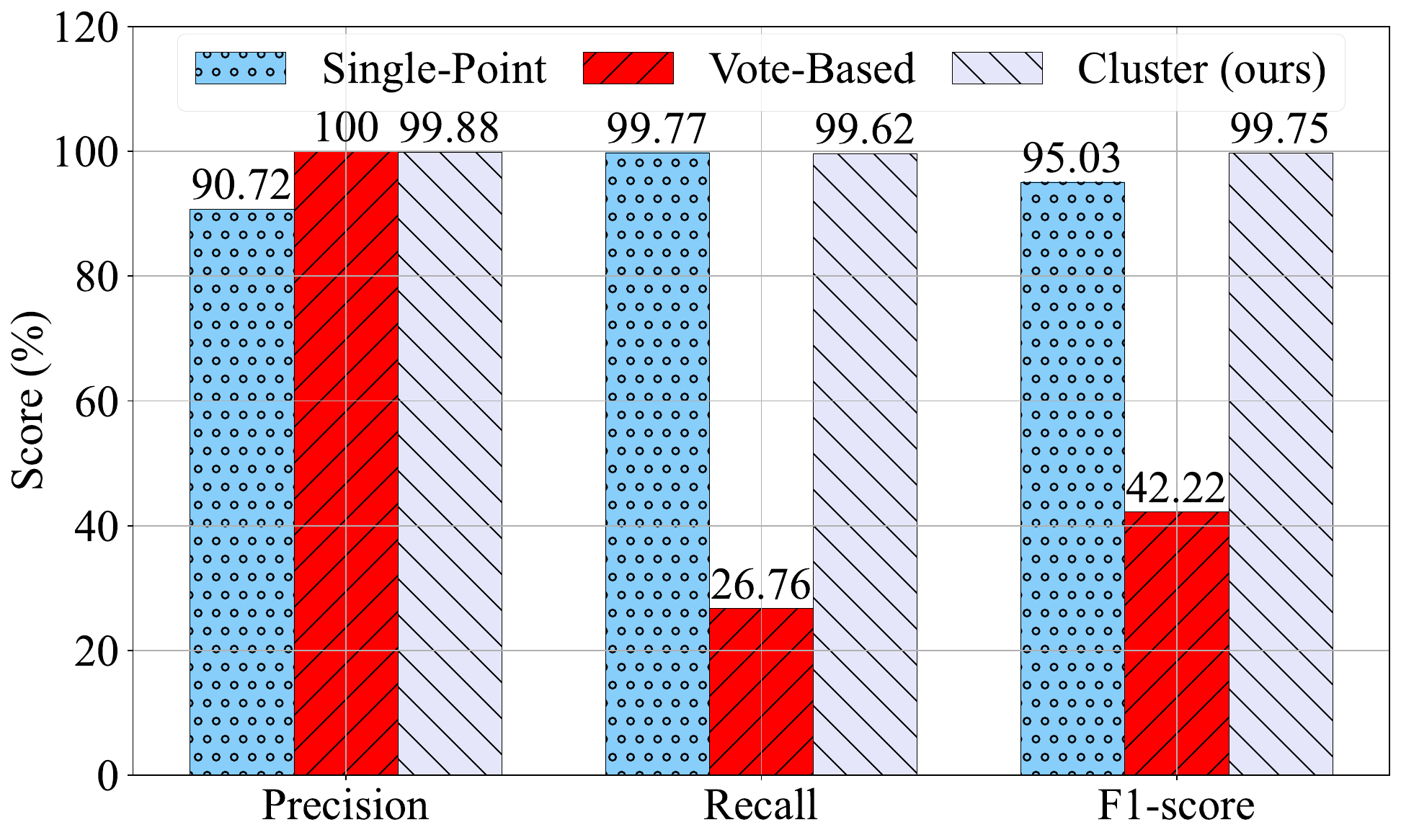}
					\label{fig: ablation-multi2single}
				\end{minipage}
			}
			\subfigure[Multi2Multi]{
				\begin{minipage}{0.47\linewidth}
					\centering
					\includegraphics[width=\textwidth]{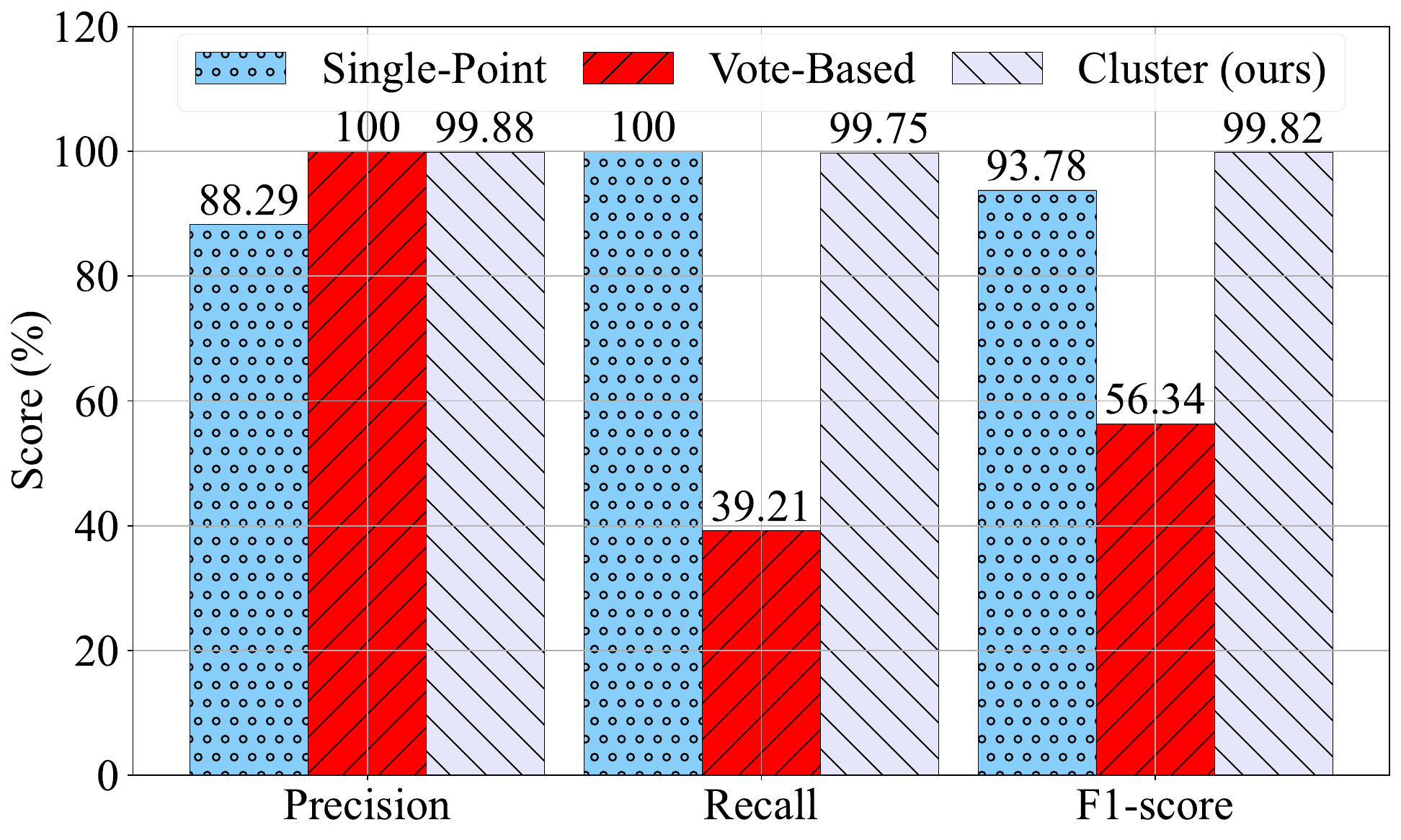}
					\label{fig: ablation-multi2multi}
				\end{minipage}
			}
			\caption{Results with(out) Cluster Classifier}
			\label{fig: ablation}
		\end{figure}
		
		As depicted in Figure~\ref{fig: ablation}, when employing Vote-Based Classification, a consistently high Precision (always 100\%) is maintained, but the Recall is low (26.76\% for Multi2Single and 39.31\% for Multi2Multi), indicating that many anomalies are missed. In contrast, using Single-Point Classification balances Precision and Recall to some extent, significantly improving the F1-Score to approximately 95\% on both datasets, surpassing the results achieved by state-of-the-art methods mentioned previously. However, with the implementation of the Cluster Classifier, the F1-Score exceeds 99\% on both datasets, demonstrating the Cluster Classifier's superior ability to utilize information from multiple nodes effectively.
		
	\end{sloppypar}
\end{document}